%% file: eval_paper.tex
\documentclass{vgtc}                          

\graphicspath{{figures/}{pictures/}{images/}{./}} 

\usepackage{times}                     
\usepackage{textgreek}
\usepackage{xcolor}

\usepackage{tabu}                      
\usepackage{booktabs}                  
\usepackage{lipsum}                    
\usepackage{mwe}                       

\usepackage{mathptmx}                  

\onlineid{2501}

\vgtccategory{Position Paper}

\vgtcinsertpkg

\title{An Evaluation-Centric Paradigm for Scientific Visualization Agents}

\author{
Kuangshi Ai\thanks{e-mail: kai@nd.edu}\\
    \scriptsize Univ. Notre Dame %
\and Haichao Miao\thanks{e-mail: miao1@llnl.gov}\\
    \scriptsize LLNL%
\and Zhimin Li\thanks{e-mail: zhimin.li@vanderbilt.edu}\\
    \scriptsize Vanderbilt Univ. %
\and Chaoli Wang\thanks{e-mail: chaoli.wang@nd.edu}\\
    \scriptsize Univ. Notre Dame %
\and Shusen Liu\thanks{e-mail: liu42@llnl.gov}\\
    \scriptsize LLNL
}

\input{body/abstract}


\teaser{
\vspace{-0.1in}
\centering
\includegraphics[width=1.0\linewidth]{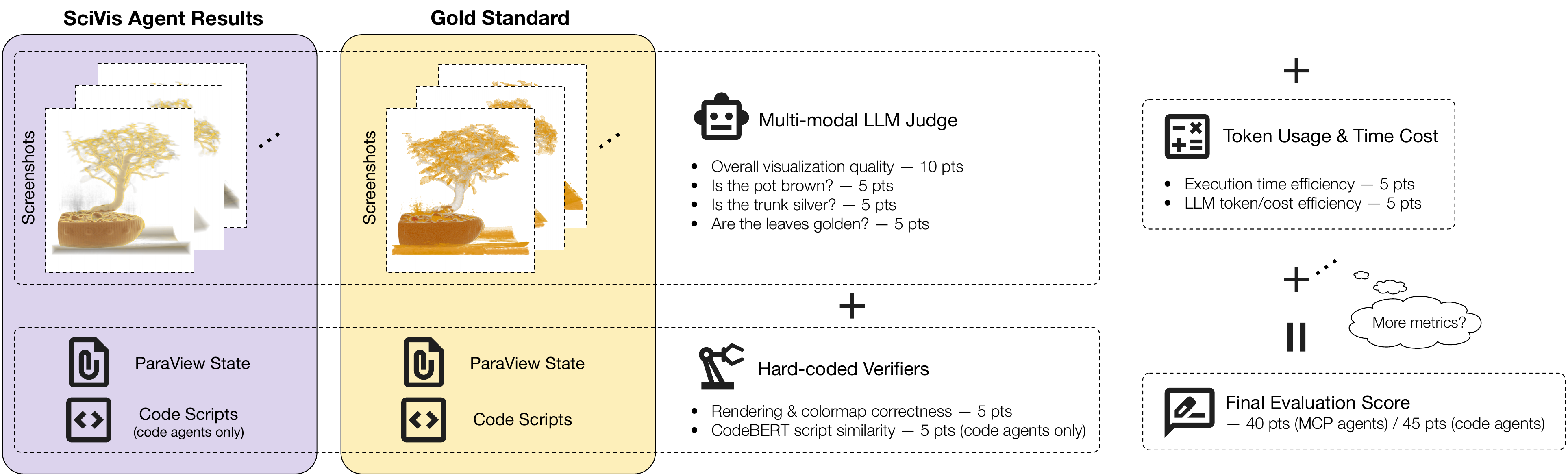}
\vspace{-0.25in}
    \caption{An example case of evaluating SciVis agents on the Bonsai dataset. Agents perform volume visualization and adjust the transfer function to achieve the target ``\emph{A potted tree with a brown pot, silver branches, and golden leaves.}" The evaluation combines: (1) a multi-modal LLM judge for visualization quality, (2) hard-coded verifiers for correctness of visualization primitives and techniques, and (3) token usage and execution time for system performance. We advocate for incorporating multifaceted metrics and developing a systematic benchmark for SciVis agent evaluation. 
}
\label{fig:teaser}
}

\begin{document}


\maketitle


\input{body/intro}
\input{body/related}
\input{body/eval_taxonomy}

\input{body/eval_effectiveness}

\input{body/call_for_collaborative_eval}

\input{body/eval_driven_design}

\input{body/discussion}



\acknowledgments{
This work was performed under the auspices of the U.S.\ Department of Energy by Lawrence Livermore National Laboratory under Contract DE-AC52-07NA27344. The work was partially supported by LLNL-LDRD (23-ERD-029, 23-SI-003), DOE ECRP (SCW1885), DOE DE-SC0023145, and the U.S.\ National Science Foundation IIS-1955395, IIS-2101696, OAC-2104158, and IIS-2401144.}

\vspace{-0.05in}
\bibliographystyle{abbrv-doi}

\bibliography{refs}
\end{document}

%% file: body/abstract.tex
\abstract{

Recent advances in multi-modal large language models (MLLMs) have enabled increasingly sophisticated autonomous visualization agents capable of translating user intentions into data visualizations. However, measuring progress and comparing different agents remains challenging, particularly in scientific visualization (SciVis), due to the absence of comprehensive, large-scale benchmarks for evaluating real-world capabilities. This position paper examines the various types of evaluation required for SciVis agents, outlines the associated challenges, provides a simple proof-of-concept evaluation example, and discusses how evaluation benchmarks can facilitate agent self-improvement. We advocate for a broader collaboration to develop a SciVis agentic evaluation benchmark that would not only assess existing capabilities but also drive innovation and stimulate future development in the field.

}

%% file: body/intro.tex

\section{Introduction}

Many breakthrough advances in machine learning (ML) and artificial intelligence (AI) began with the establishment of comprehensive and challenging benchmarks, such as ImageNet~\cite{deng2009imagenet}, that catalyzed the deep learning revolution.
These benchmarks not only provide standardized methods to evaluate and compare different techniques but also push the boundaries of what is possible with existing technology and tools.
The recent emergence of multi-modal large language models (MLLMs) has enabled a new generation of autonomous visualization agents capable of translating natural language instructions into complex scientific visualization (SciVis) results~\cite{liu2025paraview, mallick2024chatvis, ai2025nli4volvis}.
However, evaluating these agents presents a fundamental challenge: while SciVis often involves exploratory analysis with emergent insights, meaningful benchmarking requires reproducible tasks with measurable outcomes.
This evaluation gap is becoming critical as visualization agents transition from research prototypes to practical tools that scientists and engineers use.

Building on the broader taxonomy proposed by Dhanoa et al.~\cite{dhanoa2025agentic}, we adopt a more focused, practical definition for evaluation purposes.
We define a SciVis agent as: 
{\em an AI system that interprets human users' natural language intent, autonomously interacts with the SciVis pipeline to produce visualizations that meet user-specified analysis goals.}
This definition deliberately constrains the scope to enable concrete, reproducible evaluation while capturing the essential capabilities these agents must possess.
Importantly, we focus on fully autonomous execution scenarios where agents must complete tasks without additional human intervention beyond the initial instruction, allowing for consistent and repeatable benchmarking.
Current evaluation approaches for visualization agents are inadequate for SciVis tasks.
Existing benchmarks focus primarily on simple plotting tasks~\cite{chen2024viseval, yang2024matplotagent, galimzyanov2025drawing} or general data science workflows~\cite{huang2024dacode, wu2024dscode}, failing to address the unique complexity of SciVis pipelines.
Unlike basic plotting, SciVis workflows require sophisticated data transformations, diverse rendering techniques, multi-dimensional parameter mappings, and careful view selections, all of which must be applied in precise sequences to produce meaningful scientific insights.
Despite the limitations, the existing benchmark already reveals fundamental gaps in current agent capabilities, from difficulties in visual perception of visual outputs~\cite{hong2025llms, islam2024chart} to the fragility of tool-use mechanisms that underpin LLM agents~\cite{yao2024tau, qin2023toolllm}.
The absence of comprehensive evaluation frameworks not only hinders progress in the field but also makes it impossible to reliably deploy these agents in critical scientific applications where accuracy and reproducibility are paramount~\cite{report2024reproducibility, isenberg2024reproducibility}.

This position paper advocates for a fundamental shift in how we approach SciVis agent development: evaluation must become the primary design driver, not merely a validation afterthought.
We aim to catalyze broad collaboration in establishing evaluation standards that will transform SciVis agents from experimental tools into reliable scientific instruments.
In this position paper, we call for a more comprehensive evaluation benchmark that addresses multiple dimensions: task complexity (from simple parameter adjustment to complex multi-step pipelines), domain coverage (from experimental data to computational simulation), and evaluation methodology (from output quality to process efficiency).
Furthermore, we envision how such benchmarks can enable agent self-improvement through automated feedback loops, potentially leading to autonomous agent self-improvement~\cite{huautomated}.

%% file: body/related.tex
\section{Related Work}

The evaluation of AI agents has become increasingly important, with different needs for domain‑specific systems (e.g., visualization) versus general‑purpose agents. We organize prior work into (1) visualization/HCI-focused evaluations and (2) general agent benchmarks, and highlight gaps that motivate a SciVis‑specific evaluation-centric approach.

\subsection{Visualization and HCI Agent Evaluation}

\noindent\textbf{Visualization‑specific benchmarks and agents.}
Recent benchmarks show that LLMs are able to perceive basic charts but still struggle with core visualization tasks: VisEval~\cite{chen2024viseval} demonstrates failures in chart readability and generation, Drawing Pandas~\cite{galimzyanov2025drawing} exposes code executability issues, and MatPlotAgent~\cite{yang2024matplotagent} argues for dedicated evaluation beyond generic code metrics. Visualization-focused agentic pipelines have been explored, from AVA’s perception‑driven refinement to code‑generating and tool‑using SciVis agents such as ChatVis and ParaView‑MCP~\cite{liu2024ava,mallick2024chatvis,liu2025paraview}. 
Beyond output quality, several works investigate foundational capabilities: visualization literacy, and these evaluations document persistent limits in visualization understanding ~\cite{hong2025llms,shen2024visualization,islam2024chart}. Together, these limitations indicate the need for evaluation protocols that consider both the result quality and process (e.g., tool‑use) when agents rely on multi‑step SciVis pipelines.

\noindent\textbf{Human-AI collaboration.}
HCI research brings methods for assessing interaction quality and explainability in agentic workflows. Magentic‑UI emphasizes human‑in‑the‑loop evaluation; NLI4VolVis demonstrates multi‑agent, open‑vocabulary interaction for volumes; and explainability has been shown to improve task performance in human-AI teams~\cite{mozannar2025magenticui,ai2025nli4volvis,schallmoser2024explainable}. Methodological guidance for evaluating conversational assistants (e.g., modality effects and collaboration dynamics) further shows how we can assess agents that support design and analysis workflows~\cite{kuang2023collaboration,kuang2023crafting}.

\subsection{General Agent Evaluation Frameworks}

\noindent\textbf{Comprehensive agent benchmarks.}
Broader agent evaluations provide infrastructure but rarely capture SciVis's exploratory and analysis-driven demands. AgentBench and AgentBoard target multi‑turn reasoning and task success; GAIA stresses real‑world complexity; and $\tau$‑bench analyzes tool-agent-user interaction and reveals important consistency drops across trials~\cite{liu2023agentbench,chang2024agentboard,mialon2023gaia,yao2024tau}. Multi-modal and web‑task settings (e.g., VisualWebArena) and large‑scale multi-modal understanding (e.g., MMMU) underscore remaining gaps in expert‑level visual reasoning that SciVis agents must meet~\cite{koh2024visualwebarena,yue2024mmmu}. Tool‑use studies also surface fragility in API grounding~\cite{qin2023toolllm}.

\noindent\textbf{Judging, reliability, and reproducibility.}
``LLM‑as‑a‑judge" correlates reasonably with human preference but has known limits in visual grounding and stability~\cite{zheng2023judging,gu2024survey,szymanski2025limitllmjudge}. For SciVis agents, where small view/encoding changes can be semantically consequential, this motivates hybrid protocols that combine LLM judging with engine‑state verification. Finally, reproducibility remains a cross‑cutting concern in visualization and systems evaluation, reinforcing the need for transparent frameworks and  benchmarks~\cite{isenberg2024reproducibility,report2024reproducibility}, which we propose in this position paper for visualization agents. 

%% file: body/eval_taxonomy.tex
\section{A Taxonomy of SciVis Agent Evaluation}
\label{sec:eval_taxonomy}

Building on our definition of SciVis agents and the goal of making tangible progress, we propose a practical-focused taxonomy of evaluation tasks that could help drive future visualization agent development.
This taxonomy serves dual purposes: it establishes standardized metrics for comparing diverse agent architectures and provides actionable feedback for agent improvement.
We organize evaluation tasks into two primary categories: \emph{outcome-based} and \emph{process-based}.

\vspace{-0.1in}
\begin{figure}[htb]
\centering
\includegraphics[width=1.0\linewidth]{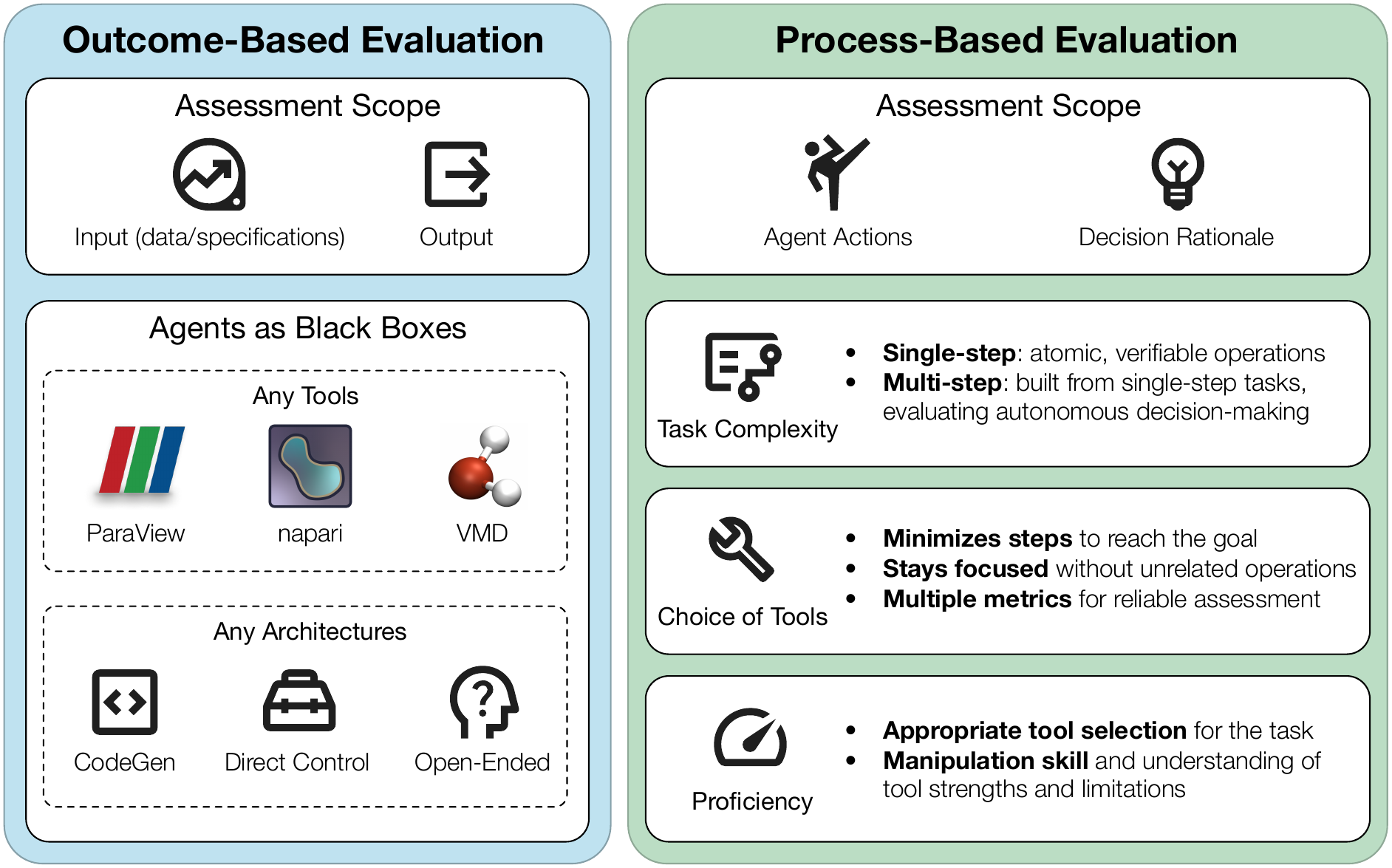}
\vspace{-0.25in}
\caption{Taxonomy of SciVis agent evaluation, organized into two perspectives: outcome-based evaluation assessing the relationship between input specifications and final outputs while treating agents as black boxes, and process-based evaluation analyzing the agent's action path, decision rationale, and intermediate behaviors.}
\label{fig:interface}
\end{figure}
\vspace{-0.1in}

\subsection{Outcome-Based Evaluation}

Outcome-based evaluation focuses exclusively on the relationship between input data/specifications and final outputs, treating the agent as a black box.
This approach is essential for ensuring broad applicability across heterogeneous agent architectures, from those that generate executable code~\cite{mallick2024chatvis} to those that directly manipulate tool interfaces~\cite{liu2025paraview} or more intelligent systems (yet to be developed) that autonomously select their approach based on task requirements.
By abstracting away implementation details, outcome-based metrics enable direct comparison between fundamentally different agent designs while maintaining focus on what ultimately matters: the quality and correctness of the visualization output. With visualization problems, one crucial challenge can arise from non-unique outcomes, i.e., different visualization results that reveal the same insights, which creates ambiguity for outcome-based evaluation. To make the agent solution specific, we can increase the constraint and condition to narrow the solution. Alternatively, we can focus on shorter, more focused tasks with no branching possibilities, or start from a predetermined intermediate result.
    
    

\subsection{Process-Based Evaluation}

Process-based evaluation examines the agent's actions and the rationale, providing insights into how solutions are achieved rather than merely what is produced.
This granular analysis is particularly valuable for identifying failure modes, understanding generalization capabilities, and guiding iterative refinement of agent architectures.
Process-based evaluation comes with added complexity. We formulate the following subcategories with the following focus.

\noindent\textbf{Task complexity} naturally divides process-based evaluations, i.e., single-step vs.\ multi-step tasks. Single-step tasks evaluate atomic operations such as loading a dataset and applying a specific filter. 
Multi-step tasks consist of interdependent single-step tasks spanning dozens or even hundreds of steps, 
potentially involving backtracking and iterative refinement. 
While multi-step tasks allow diverse exploration trajectories, each constituent single-step task should remain verifiable and consistent in its objectives, e.g., exemplified by the VeriGUI dataset~\cite{liu2025verigui}.
    
\noindent\textbf{Choice of tools} represents another critical dimension for the visualization pipeline, as SciVis encompasses a broad ecosystem of specialized software/packages. Evaluation can be divided based on targeted tools such as ParaView for general SciVis, napari for biomedical imaging, VMD for molecular dynamics, etc. Advanced agents might demonstrate meta-capabilities by autonomously selecting the most appropriate tool for a given task, requiring not only the ability to manipulate and use different tools but also an understanding of each tool's strengths and limitations.

\noindent\textbf{Proficiency} is another important dimension for process evaluation. Does the agent use more steps than strictly required? Does it do unrelated tasks concerning the stated goals while stumbling upon the correct solution? Regarding practical measurement, we can rely indirectly on token usage and time or step length.





%% file: body/eval_effectiveness.tex
\section{Effectiveness of Agent Evaluation}
\label{sec:eval_effectiveness}







Even the most comprehensive benchmark would be useless if the evaluations are not trustworthy, i.e., failing to reflect a given agent's full capabilities and shortcomings accurately.  
Effectiveness of the evaluation can be examined through three complementary lenses: \emph{accuracy}, which concerns the reliability of individual evaluation results; \emph{coverage}, which indicate how much of the potential real-world usage scenarios are covered by the benchmark; and \emph{cost-effectiveness}, which convey the need to strike a balance between the amount of computational and human efforts and achieving good accuracy and coverage.

\noindent\textbf{Accuracy} requires reducing uncertainty and ensuring robust evaluation signals. One viable approach for assessing visualization quality is to use MLLM judges, which have shown strong alignment with human preference~\cite{zheng2023judging,gu2024survey}.
However, recent studies~\cite{szymanski2025limitllmjudge, berger2024visjudge} reveal that these models still face notable limitations in visual perception and grounding despite their promise. They may overlook subtle visual encodings, misinterpret spatial relationships, or conflate stylistic variation with semantic differences, and their judgments can drift with prompt phrasing or image presentation order.
For greater reliability, automated verification against the visualization engine’s internal state can be used (process-based evaluation). For example, a case-specific Python script can confirm that a ParaView isosurface has been generated at the correct value and colored appropriately. Quantitative checks can be added for code-generating agents such as ChatVis by comparing generated scripts with gold-standard reference scripts and validating their execution outcomes. Human evaluation, though costly, may still be needed for ambiguous or high-stakes cases.


\noindent\textbf{Coverage} concerns whether the evaluation spans the full range of SciVis tasks and interaction patterns. Test design should begin with representative user intents mapped to diverse techniques (e.g., volume rendering, streamline tracing, isosurface extraction). An outcome-based evaluation specifies only the dataset and task description, without constraining how agents achieve the goal. This allows fair evaluation of agents with varying capabilities, whether they generate code to interact with the visualization engine or directly invoke high-level tools. Ensuring evaluation coverage benefits from both top-down alignment with a taxonomy of visualization tasks and bottom-up analysis of which visualization primitives, techniques, and interaction modalities are exercised. This dual perspective helps identify gaps and keeps the benchmark representative of real-world use cases.

\noindent\textbf{Cost-effectiveness} addresses two key practical constraints in SciVis agent evaluation. First, defining ground truth for exploratory SciVis tasks is inherently challenging: unlike deterministic operations, these tasks often allow multiple equally valid visualizations, viewpoints, or parameter settings. This ambiguity complicates the creation of automated verifiers and can require costly human judgment to establish fair scoring criteria. Second, running comprehensive evaluations—spanning diverse datasets, visualization techniques, and agent configurations—demands substantial computational resources. The need to repeatedly launch visualization engines, process large scientific datasets, and execute complex visualization pipelines can lead to prohibitive runtime and monetary costs. Benchmarks must strike a balance, delivering actionable and representative evaluation while minimizing overhead to support rapid, iterative development cycles.

%% file: body/call_for_collaborative_eval.tex
\section{An Example for SciVis Agent Evaluation}
\label{sec:eval_collab}

To ground the discussion in a concrete setting, we outline an example benchmark for SciVis agents. The aim is to show how \emph{outcome quality}, \emph{process verification}, and \emph{system efficiency} can be combined into a unified evaluation protocol.

\subsection{Framework Design}

We present this setup as an illustrative example of how one might design an evaluation protocol for SciVis agents, aligned with the taxonomy in Section~\ref{sec:eval_taxonomy}. Tasks are decomposed into smaller, controllable checkpoints to pinpoint points of failure, and agents operate in a controlled sandbox via either model context protocol (MCP)~\cite{anthropic2024mcp} or direct code execution.

For \textbf{outcome quality}, the focus is on whether the final visualization meets the intended goals in terms of accuracy, semantic correctness, and interpretability. Factors such as colormap selection, viewpoint, and use of appropriate visualization primitives are assessed. We employ instruction-tuned multi-modal LLM judges aligned with human preferences in our implementation. These models are prompted with domain-specific evaluation criteria, the ground-truth visualization, and the agent’s output, then asked to assign quality scores.

For \textbf{process verification}, the emphasis is on whether the agent’s intermediate actions and applied techniques satisfy explicit task requirements. This includes verifying the correct use of visualization primitives (e.g., isosurfaces) and techniques (e.g., volume rendering) via case-specific hard-coded verifiers that inspect the visualization engine’s internal state. For code-generating agents, additional checks compare the generated scripts to gold-standard references and validate their execution outcomes.

For \textbf{system efficiency}, we track runtime, token usage, and monetary cost for each run. These measures complement accuracy-based metrics, providing insight into scalability, cost-effectiveness, and the real-world deployability of agentic visualization systems.

\subsection{Illustrative Case Study: Bonsai Volume Rendering}

As a concrete illustration of the proposed SciVis agent evaluation framework, we consider a volume rendering task on the Bonsai dataset using ParaView as the visualization pipeline. Two agents are evaluated: ChatVis~\cite{mallick2024chatvis}, which generates Python scripts to interact with ParaView's native APIs, and ParaView-MCP~\cite{liu2025paraview}, which operates through an MCP server, a higher level of abstraction over the raw APIs. In this example, both agents use models from the GPT series, i.e., GPT-5, GPT-4.1, and GPT-4o, as their backbone LLM. Each experiment is repeated 10 times for statistical robustness.
The agents are instructed to load the Bonsai dataset with given parameters, perform volume rendering, and adjust the transfer function to achieve the target visualization: ``\emph{A potted tree with a brown pot, silver branches, and golden leaves.}" The resulting ParaView state is saved for subsequent evaluation. 

Upon task completion, overall visualization quality is assessed using an instruction-tuned multi-modal LLM judge (e.g., GPT-4o),
presented with both the ground-truth images and the agent-generated results. The judge evaluates outputs against explicit criteria: whether the overall goal is met, whether the pot is brown, the branches are silver, and the leaves are gold. These scores form part of the final evaluation metric.

To enhance robustness of the assessment, the LLM-based evaluation is supplemented with hard-coded verification scripts executed via \texttt{pvpython}. The saved ParaView state is reloaded to confirm the correct volume rendering configuration and accurate colormap settings. For code-generating agents such as ChatVis, we additionally compute the CodeBERT-based~\cite{feng2020codebert} similarity between the generated and gold-standard reference scripts. While these case-specific checks substantially improve reliability, they require additional manual effort to design and maintain.
Performance metrics, including token usage, monetary cost, and task completion time, are recorded as they directly reflect user-perceived latency and the practical feasibility of deploying such agents. Each metric is assigned a point value, and the sum of these points constitutes the final evaluation score (see Figure~\ref{fig:teaser}).
Table~\ref{tab:scivis_agent_results} shows that while the MCP-based agent delivers stable, high-quality results, its reliance on complex toolchains leads to high latency, limiting real-world deployment.
In contrast, ChatVis—lacking vision capabilities and generating code on the fly—typically completes tasks more quickly but incurs increased token usage and reduced visualization quality.

While the results focus on GPT series models, we also evaluated other model families such as Claude, LLaMA, and Qwen, and observed significant variation in SciVis task performance across models. When paired with highly abstracted tool environments like ParaView-MCP, smaller language models (SLMs) often achieve comparable visualization quality with lower latency and reduced cost. In such settings, strong reasoning ability is less critical, allowing SLMs to complete visualization tasks effectively~\cite{belcak2025small}. However, the advanced visual understanding capabilities of larger models like GPT-5 indeed lead to improved visualization outcomes. Given the absence of a systematic evaluation protocol for SciVis agents, we advocate for creating a comprehensive benchmark to guide future research and development.

\vspace{-0.1in}
\begin{table}[ht]
\centering
\caption{Evaluation results of two SciVis agents on the Bonsai task using models from the GPT-series as a backbone. Each experiment is repeated 10 times. The mean and variance of token usage and time cost are reported, along with the best SciVis evaluation score for each setting. SR denotes the success rate. Results were obtained on Sep 17, 2025 using the OpenAI API.}
\resizebox{\columnwidth}{!}{
\begin{tabular}{lllllll}
\textbf{agent} & \textbf{model} & \textbf{I/O tokens} & \textbf{avg cost} & \textbf{time (s)} & \textbf{SR} & \textbf{score} \\
\hline
MCP-based & GPT-5   & 220 $\pm$ 0 / 838 $\pm$ 203 & \$0.0087 & 301.7 $\pm$ 32.3 & 10/10 & 27/40 \\
ChatVis   & GPT-5   & 2430 $\pm$ 847 / 2994 $\pm$ 956 & \$0.0330 & 158.9 $\pm$ 29.9 & 10/10 & 25/45 \\
MCP-based & GPT-4.1 & 220 $\pm$ 0 / 1460 $\pm$ 210 & \$0.0121 & 49.3 $\pm$ 8.0 & 10/10 & 21/40 \\
ChatVis   & GPT-4.1 & 638 $\pm$ 555 / 1217 $\pm$ 530 & \$0.0110 & 24.0 $\pm$ 5.7 & 10/10 & 23/45 \\
MCP-based & GPT-4o  & 220 $\pm$ 0 / 908 $\pm$ 109 & \$0.0239 & 41.7 $\pm$ 14.2 & 10/10 & 23/40 \\
ChatVis   & GPT-4o  & 1945 $\pm$ 753 / 1909 $\pm$ 672 & \$0.0240 & 38.4 $\pm$ 9.4 & 7/10 & 24/45 \\
\end{tabular}
}
\label{tab:scivis_agent_results}
\end{table}
\vspace{-0.1in}

%% file: body/eval_driven_design.tex
\section{Evaluation-Driven Agent Design}
\label{sec:eval_driven_design}


Developing SciVis agents requires integrating numerous tools and libraries, each demanding substantial engineering effort for LLM-based control.
Rather than tackling this complexity through traditional development, where evaluation follows implementation, we propose inverting this relationship: comprehensive evaluation benchmarks can drive the entire agent design process.

Drawing inspiration from test-driven development, evaluation-driven design uses benchmarks as both specification and scaffold for incremental agent development.
Developers can build capabilities progressively, validating single-step operations before advancing to complex multi-step workflows, transforming an overwhelming engineering challenge into manageable iterations.
Each evaluation target provides concrete guidance, accelerating development while ensuring robust functionality.
Recent work on self-evolving AI agents~\cite{fang2025selfevolveagent}, with meta agents for automated agent design~\cite{huautomated} as a representative approach, has demonstrated the viability of a similar approach.
Process-based evaluations identify specific reasoning failures and inefficient tool selections, while outcome-based evaluations provide clear optimization targets.
Meta-agents can analyze these results to modify code or prompts automatically.

The relationship between evaluation and agent development is symbiotic, as agents grow more capable, benchmarks expand to challenge new abilities. Moreover, when evaluations become more comprehensive, they reveal hidden failure modes and guide practical improvements.
This co-evolution ensures benchmarks remain relevant while agents develop robust, generalizable capabilities rather than overfitting to artificial metrics.
Effective evaluation-driven design requires benchmarks that provide granular, actionable feedback rather than binary success/failure signals.
They must span from deterministic single-step operations to open-ended exploratory tasks, and execute efficiently to enable rapid iteration cycles.
The extensive evaluation suite we advocate for thus serves not merely as a measurement instrument but as a development accelerator, fundamentally transforming how we construct SciVis agents.
By making evaluation the primary driver rather than a validation afterthought, we can build more capable and reliable systems while significantly reducing development time and effort.

%% file: body/discussion.tex
\section{Concluding Remarks}



While our outlined evaluation framework provides a structured path toward evaluating and improving SciVis agents, we acknowledge important limitations.
By restricting evaluation to fully autonomous scenarios without human interaction beyond initial input, we exclude the critical domain of human-AI collaboration.
Evaluating human-AI interaction—with its variability in user expertise and communication styles—represents a distinct research area beyond our current scope. However, simulated multi-turn evaluation approaches~\cite{liu2023agentbench} offer promising directions for capturing some interactive dynamics without direct evaluation with human users.

The deployment of autonomous visualization agents also raises critical safety considerations that our framework must address.
Agents with direct access to tools and code execution could potentially corrupt data and consume excessive computational resources.
We advocate for sandboxed evaluation environments that isolate agent execution from production systems, similar to approaches used in general agent benchmarks~\cite{bonatti2024windows, zhou2024webarena}.
Furthermore, evaluation-driven self-evolving agents present unique risks—automated refinement loops could amplify harmful behaviors or exploit evaluation metrics in unintended ways.
These safety concerns necessitate careful monitoring, bounded optimization objectives, and human oversight checkpoints even in autonomous evaluation scenarios, ensuring that the pursuit of improved benchmark performance aligns with safe and reliable scientific practice.

Creating comprehensive evaluation benchmarks for SciVis agents exceeds the capacity of any single research group.
The diversity of scientific domains, visualization tools, and use cases demands broad collaboration between visualization researchers, domain scientists, AI practitioners, and tool developers.
\textbf{This position paper serves as an open invitation for collaboration on building this evaluation benchmark as a community.} 
Such partnerships would ensure benchmarks reflect genuine scientific needs rather than artificial constructs, while distributing the substantial effort required to create, validate, and maintain evaluation suites.
The evaluation suites can and should be extended or expanded as overall objectives evolve with new advancements.
Future directions include evaluation of collaborative multi-agent systems where specialized agents coordinate on complex visualization tasks, assessment of integration of domain-specific knowledge into evaluation metrics for more nuanced judgment of scientific insights, and development of benchmarks for creative visualization approaches beyond established techniques.
Establishing rigorous evaluation frameworks today lays the foundation for SciVis agents that genuinely augment human scientific inquiry, transforming how researchers explore and understand complex data.